\documentclass{article}
\begin{document}
\title{The Geometry of Quantum Mechanics}
\author{Jos\'e M. Isidro\\
Department of Theoretical Physics,\\ 
1 Keble Road, 
Oxford OX1 3NP, UK.\\ 
{\tt isidro@thphys.ox.ac.uk}}
\maketitle

\begin{abstract}

A recent notion in theoretical physics is that not all quantum theories arise 
from quantising a classical system. Also, a given quantum model may possess more 
than just one classical limit. These facts find strong evidence in string 
duality and M--theory, and it has been suggested that they should also 
have a counterpart in quantum mechanics. In view of these developments 
we propose {\it dequantisation}, a mechanism to render a quantum theory 
classical. Specifically, we present a geometric procedure to {\it dequantise}
a given quantum mechanics (regardless of its classical origin, if any) 
to possibly different classical limits, whose quantisation gives back 
the original quantum theory. The standard classical limit $\hbar\to 0$ 
arises as a particular case of our approach.


2001 Pacs codes: 03.65.-w, 03.65.Ca


Preprint no. OUCP-01-51P

\end{abstract}

\tableofcontents

\section{Introduction}\label{intro}

\subsection{Motivation}\label{moti}

Appraoching quantum mechanics from a geometric viewpoint is a very interesting 
topic. The goal is a geometrisation of quantum mechanics \cite{ASHTEKAR},
similar in spirit to that of classical mechanics \cite{ARNOLD, RATIU}. 
Beyond this similarity, however, there are numerous deep reasons. 
One of them is motivated in string duality and M--theory \cite{SCHWARZ, VAFA}. 
In plain words, we are confronted with the fact that not all quantum theories 
arise from quantising a classical system. Also, a given quantum model may possess 
more than just one classical limit. These two facts are in sharp contrast with our 
current understanding of quantum mechanics. While it is true that these two phenomena 
originally arise in the theories of strings and branes \cite{KAKU}, 
some authors \cite{VAFA} have expressed the opinion that they should 
somehow be reflected at the fundamental level of quantum mechanics as well. 
Let us describe the general setup.
 
Quantisation may be understood as a prescription to construct a quantum theory from a given 
classical theory. As such, it is far from being unique. Beyond canonical quantisation 
and functional integrals, a number of  different, often complementary approaches 
to quantisation are known, each one of them exploiting different aspects of the underlying 
classical theory. For example, geometric quantisation \cite{KOSTANT, SOURIAU, 
SNIA, WOODHOUSE} relies on the geometric properties of classical mechanics. 
Systems whose classical phase space 
${\cal C}$ is a K\"ahler manifold can be quantised as in refs. \cite{BEREZIN, SCHLICHENMAIER, RT}. 
If ${\cal C}$ is just a Poisson manifold, then the  approach of ref. \cite{KONTSEVICH}, 
based on deformation quantisation \cite{STERNHEIMER, ZACHOS}, can always  be applied. 
A path--integral counterpart to these mathematical techniques has been developed 
in ref. \cite{CATFELDER}.

A common feature to these approaches is the fact that they all take a classical mechanics 
as a starting point. Thus the classical limit is {\it a fortiori} unique: it reduces to 
letting $\hbar\to 0$. If we want to allow for the existence of more than one classical limit, 
we are led to considering a quantum mechanics that is not based, at least primarily, 
on the the quantisation of a given classical dynamics. In such an approach 
one would not take first a classical model and then quantise it. Rather, 
quantum mechanics itself would be the starting point: a parent quantum theory
would give rise, in a certain limit, to a classical theory. 
If there are several different ways of taking this limit, 
then there will be several different classical limits. 

\subsection{Summary}\label{suma}

This article puts forward a geometric proposal by which quantum mechanics 
can be rendered classical, or {\it dequantised}, in more than one way, 
thus yielding different classical limits. Under {\it dequantisation} 
we understand the following. 

Assume that classical phase space ${\cal C}$ is ${\bf R}^{2n}$. Then,
starting from the quantum phase space ${\cal Q}$ of standard quantum mechanics 
\cite{ASHTEKAR}, the usual classical limit $\hbar\to 0$ is obtained as the quotient 
of ${\cal Q}$ by a certain equivalence relation $\sim$, {\it i.e.}
${\cal Q}/\sim\,={\bf R}^{2n}$,
and we have a trivial fibre bundle ${\cal Q}\rightarrow {\bf R}^{2n}$. 
We will construct classical phase spaces ${\cal Q}/G={\cal C}$, 
where $G$ is a Lie group acting on ${\cal Q}$, and ${\cal Q}\rightarrow {\cal C}$ 
will be a (not necessarily trivial) $G$--bundle. The associated vector 
bundle will have ${\cal H}$, the Hilbert space of quantum states, as its 
typical fibre. In order to qualify as a classical 
phase space, ${\cal C}$ must be a symplectic manifold whose quantisation must give back 
the original quantum theory on ${\cal Q}$. Different choices for $G$ will give rise to 
different classical limits. 

\subsection{Outline}\label{outl}

This article is organised as follows. Section \ref{ast} summarises the standard Hilbert space 
formulation of quantum mechanics, following the geometric presentation of ref. \cite{ASHTEKAR}. 
We will recall how the standard classical limit $\hbar \to 0$ is taken. In this analysis, 
a natural mechanism will arise that will allow more than one classical limit to exist. 
This is presented in section \ref{claslim}. We illustrate our technique with some 
specific examples in section \ref{exam}, where one given quantum mechanics is explicitly 
{\it dequantised}. The physical implications of our proposal are discussed in section \ref{discu}. 
Some technical mathematical aspects of our construction are elucidated in section \ref{mathtech}.
 
\section{A geometric approach to quantum mechanics}\label{ast}

For later purposes let us briefly summarise the geometric approach to quantum mechanics 
presented in ref. \cite{ASHTEKAR}. Throughout this section our use of the terms {\it classical} 
and {\it quantum} will be the standard one \cite{GALINDO}.

\subsection{The Hilbert space as a K\"ahler manifold}\label{hika}

The starting point is an infinite--dimensional, complex, separable Hilbert space of quantum states, 
${\cal H}$, that is most conveniently viewed as a real vector space equipped with a complex 
structure $J$. Correspondingly, the Hermitian inner product can be decomposed into real and 
imaginary parts,
\begin{equation}
\langle\phi,\psi\rangle = g(\phi,\psi) + {\rm i}\omega(\phi,\psi),
\label{dec}
\end{equation}
with $g$ a positive--definite, real scalar product and $\omega$ a symplectic form.
The metric $g$, the symplectic form $\omega$ and the complex structure $J$ are related as
\begin{equation}
g(\phi,\psi)=\omega(\phi,J\psi),
\label{kah}
\end{equation}
which means that the triple $(J, g, \omega)$ endows the Hilbert space 
${\cal H}$ with the structure of a K\"ahler space \cite{ARNOLD}. 

Thus any Hilbert space naturally gives rise to a symplectic manifold:
it is the {\it quantum} phase space ${\cal Q}$, or the space of rays in ${\cal H}$. 
Let $\omega_{\cal Q}$ denote the restriction of $\omega$ to ${\cal Q}$.
On ${\cal Q}$, the inverse of $\omega_{\cal Q}$ can be used to define Poisson brackets 
and Hamiltonian vector fields. This is done as follows.

Any function $f_{\cal C}:{\cal C}\to {\bf R}$ defined on classical phase 
space ${\cal C}$ has associated a self--adjoint quantum observable $F$
on ${\cal H}$. The latter gives rise to a quantum function $f_{\cal Q}:{\cal Q}\to {\bf R}$
on quantum phase space ${\cal Q}$, defined as the expectation value of the operator $F$:
\begin{equation}
f_{\cal Q}(\psi)=\langle\psi,F\psi\rangle.
\label{func}
\end{equation} 
Now every function $f:{\cal Q}\to {\bf R}$ defines a Hamiltonian vector field 
$X_f$ through the equation \cite{RATIU}
\begin{equation}
i_{X_f} \omega_{\cal Q} = {\rm d} f.
\label{hamvf}
\end{equation}
In this way the Poisson bracket $\{\,,\}_{\cal Q}$ on ${\cal Q}$ is defined 
by \cite{RATIU}
\begin{equation}
\{f_{\cal Q}, g_{\cal Q}\}_{\cal Q} = \omega_{\cal Q} (X_f, X_g).
\label{pois}
\end{equation} 
Let us now consider the classical coordinate and momentum  
functions $q^j_{\cal C}$ and $p^k_{\cal C}$ satisfying the 
canonical Poisson brackets on ${\cal C}$. Through the above construction one arrives at 
the quantum coordinate and momentum functions $q^j_{\cal Q}$ and $p^k_{\cal Q}$
satisfying the canonical Poisson brackets on ${\cal Q}$
\begin{equation}
\{q^j_{\cal Q}, p^k_{\cal Q}\}_{\cal Q}=\delta^{jk},\qquad
\{q^j_{\cal Q}, q^k_{\cal Q}\}_{\cal Q}=0=\{p^j_{\cal Q}, p^k_{\cal Q}\}_{\cal Q}.
\end{equation}
It turns out that Hamilton's canonical equations of motion on ${\cal Q}$ are equivalent 
to Schr\"odinger's  wave equation, while the Riemannian metric $g$ accounts for properties 
such as the measurement process and Heisenberg's uncertainty relations.

We are thus dealing with two phase spaces, that we denote ${\cal C}$ (for {\it classical}\/) 
and ${\cal Q}$ (for {\it quantum}). ${\cal Q}$ is always infinite--dimensional, 
as it derives from an infinite--dimensional Hilbert space. 
On the contrary, ${\cal C}$ may well be finite--dimensional. 
Furthermore, while both ${\cal C}$ and ${\cal Q}$ are symplectic manifolds, 
the latter is always K\"ahler, while the former need not be K\"ahler.

Two questions arise naturally. First, what is the geometric relation between ${\cal C}$ 
and ${\cal Q}$ as manifolds? Second, how are ${\cal C}$ and ${\cal Q}$ related as 
{\it symplectic} manifolds, {\it i.e.}, how are their respective symplectic forms 
$\omega_{\cal C}$ and $\omega_{\cal Q}$ related? When ${\cal C}={\bf R}^{2n}$, 
the answer is provided in ref. \cite{ASHTEKAR} and summarised below.

\subsection{Quantum phase space as a fibre bundle over classical phase space}\label{fibover}

For a classical system with $n$ degrees of freedom, let us collectively 
denote by $f_r$, $r=1,\ldots, 2n$, the quantum coordinate 
and momentum functions $q^j_{\cal Q}$ and $p^k_{\cal Q}$. 
We define an equivalence relation on ${\cal Q}$ as
\begin{equation}
x_1\sim x_2\qquad {\rm iff} \qquad f_r(x_1)=f_r(x_2)\quad \forall r.
\label{equival}
\end{equation}
Through this equivalence relation, the quantum phase space ${\cal Q}$ becomes 
a trivial fibre bundle with fibre ${\cal H}$ over the classical phase space 
${\bf R}^{2n}$:
\begin{equation}
{\cal Q}\longrightarrow {\cal Q}/\sim\;={\bf R}^{2n}.
\label{mod}
\end{equation} 

\subsection{Relation between the classical and the quantum symplectic forms}\label{hori}

A tangent vector $v\in T_x{\cal Q}$ is said vertical at $x\in{\cal Q}$ 
if $v(f_r)=0$ $\forall r$. Therefore the vertical directions are those 
in which the quantum coordinate and momentum functions assume constant values.
Equivalently, the vertical subspace ${\cal V}_x$ at $x\in{\cal Q}$ may be defined as
\begin{equation}
{\cal V}_x = \{v\in T_x{\cal Q}: \omega_{\cal Q}\left(X_{f_r}(x), 
v\right)=0\; \forall r\}.
\label{vvert}
\end{equation}
Let ${\cal V}_x^{\perp}$ denote the $\omega_{\cal Q}$--orthogonal 
complement of the vertical subspace at $x\in {\cal Q}$. Each tangent space 
splits as the direct sum
\begin{equation}
T_x{\cal Q}={\cal V}_x\oplus {\cal V}_x^{\perp},
\label{split}
\end{equation}
and the tangent vectors that lie in ${\cal V}_x^{\perp}$ are said horizontal at $x$. 
It turns out that the quantum states lying on a horizontal cross section of the 
bundle (\ref{mod}) are precisely the generalised coherent states of refs. 
\cite{COHST, PERELOMOV}.

Now, if $u$ and $v$ are vectors on ${\cal C}={\bf R}^{2n}$, denote by $u^h$ and $v^h$ 
their horizontal lifts to ${\cal Q}$. Then the classical symplectic 
structure $\omega_{\cal C}$ is related to its quantum 
counterpart $\omega_{\cal Q}$ through 
\begin{equation}
\omega_{\cal C}(u,v)=\omega_{\cal Q}(u^h, v^h),
\label{lift}
\end{equation}
{\it i.e.}, $\omega_{\cal C}$ is the horizontal part of $\omega_{\cal Q}$. 

\section{Taking a classical limit}\label{claslim}

The geometric presentation summarised in section \ref{ast} makes it clear 
that the quantum theory contains all the information about the classical theory. 
In this sense, as explained in section \ref{intro}, we should think of 
quantum mechanics as being prior to classical mechanics. Rather than 
quantising a classical theory, rendering quantum mechanics classical, 
or {\it dequantising}\/ it, appears to be the key issue. How can one 
{\it dequantise}\/? 

\subsection{Symplectic reduction}\label{rol}

Our primary concern will be to obtain a classical {\it symplectic} manifold 
$({\cal C},\omega_{\cal C})$ from its quantum counterpart $({\cal Q}, \omega_{\cal Q})$, 
in such a way that the quantisation of $({\cal C},\omega_{\cal C})$ will reproduce $({\cal Q}, 
\omega_{\cal Q})$ as a {\it symplectic} manifold,  regardless of the Riemannian metric 
$g_{\cal C}$ on ${\cal C}$, if any. In principle, {\it dequantisation} may be thought of 
as the symplectic reduction from $({\cal Q}, \omega_{\cal Q})$ to a symplectic submanifold 
$({\cal C},\omega_{\cal C})$; a more general definition will be given in 
section \ref{foli}. In having ${\cal C}$ as a reduced symplectic 
manifold of ${\cal Q}$ we are assured that the quantisation of ${\cal C}$ 
reproduces ${\cal Q}$. See refs. \cite{RATIU, WEINSTEIN} 
for a treatment of symplectic reduction.

We do not require the metric $g_{\cal Q}$ on ${\cal Q}$ to descend to a metric $g_{\cal C}$ 
on ${\cal C}$. Disregarding the metric $g_{\cal C}$ is justified, as the metric $g_{\cal Q}$ 
of eqn. (\ref{dec}) can always be obtained from the symplectic form $\omega_{\cal Q}$ 
through the K\"ahler condition (\ref{kah}).
 
On the contrary, the symplectic structure is an essential ingredient to keep in the passage 
from quantum to classical, as classical phase space is always symplectic. 
In what follows we will consider symplectic structures as in refs. \cite{GS, MUELLER} but, 
more generally, one could relax ${\cal C}$ to be a Poisson manifold. 

\subsection{Reduction via fibre bundles}\label{diffib}

A useful approach to symplectic reduction is via fibre bundles. 
When ${\cal C}={\bf R}^{2n}$, the classical limit arises in ref. \cite{ASHTEKAR} 
as the base space of a trivial fibre bundle with fibre ${\cal H}$ and total space ${\cal Q}$. 
This suggests considering fibre bundles ${\cal Q}\rightarrow {\cal C}$, with fibre ${\cal H}$ 
and total space ${\cal Q}$, over some other finite--dimensional base manifold ${\cal C}$.
If the classical phase space ${\cal C}$ so obtained is a symplectic manifold 
whose  quantisation reproduces the initial quantum theory on ${\cal Q}$, 
then associated with that fibre bundle there is one classical limit.

Let us first examine trivial fibre bundles. The equivalence relation of section \ref{fibover} 
is singled out because it is well suited to obtain the standard coherent states of refs. 
\cite{COHST, PERELOMOV}. We will see in section \ref{triexam} one particular example 
of a certain group $G$ acting on ${\cal Q}$ such that ${\cal Q}/G={\cal C}$ coincides 
with the result of taking the standard classical limit $\hbar\to 0$. The procedure 
of section \ref{triexam} is in fact quite general in order to replace equivalence relations 
with group actions.

Nontrivial fibre bundles may also be considered. They provide a realisation of the statement 
presented in ref. \cite{NOS}, to the effect that one can always choose local 
coordinates on classical phase space, in terms of which quantisation 
becomes a local expansion in powers of ${\hbar}$ around a certain local 
vacuum.
This expansion is local in nature: it does not hold globally on classical phase space
when the fibre bundle is nontrivial. In this sense, quantisation is 
mathematically reminiscent of the local triviality property satisfied by every 
fibre bundle \cite{STEENROD} while, physically, it is reminiscent of the 
equivalence principle of general relativity \cite{WEINBERG}.

\subsection{Definition of {\it dequantisation}}\label{foli}

For our purposes, {\it dequantisation} will mean the following.
Let $G$ a Lie group acting on ${\cal Q}$. Modding out by the action of $G$
we will construct principal $G$--bundles
\begin{equation}
{\cal Q}\longrightarrow {\cal Q}/G={\cal C}
\label{pri}
\end{equation}
over finite--dimensional symplectic manifolds ${\cal C}$.  We require 
the associated vector bundle to have ${\cal H}$ as its fibre. 
Moreover the lift of $\omega_{\cal C}$ to ${\cal Q}$ 
must equal $\omega_{\cal Q}$. 

Eqn. (\ref{lift}) expressed the property that, when ${\cal C}={\bf R}^{2n}$, 
$\omega_{\cal C}$ was simply the horizontal part of $\omega_{\cal Q}$.
Horizontality was closely related to coherence. 
Here we have no notion of horizontality because {\it any} 
$\omega_{\cal C}$ will work, provided its lift to ${\cal Q}$ 
equal $\omega_{\cal Q}$ (as is the case, {\it e.g.}, in symplectic reduction).
In general, the best we can do is to find local canonical coordinates on ${\cal C}$ 
in terms of which 
\begin{equation}
\omega_{\cal C}={\rm d} p_k \wedge {\rm d} q^k.
\label{cano}
\end{equation}
With respect to these local coordinates, local coherent states $|z_k\rangle$ 
can be defined simply as eigenvectors of the local annihilation operator 
$a_k=Q^k+{\rm i} 
P_k$, where $Q^k$ and $P_k$ are the quantum observables corresponding to 
$q^k$ and $p_k$. How do $Q^k$ and $P_k$ {\it dequantise} to $q^k$ and $p_k$?

\subsection{Classical functions from quantum observables}\label{clafunct}

When {\it dequantising}, instead of having classical functions 
$f_{\cal C}:{\cal C}\to{\bf R}$ 
to turn into quantum observables $F$, we have quantum observables $F$ out of which 
we would like to obtain classical functions. We can use eqn. (\ref{func}) in order to 
define the quantum function $f_{\cal Q}:{\cal Q}\to{\bf R}$ corresponding 
to the observable $F$. Now, in the examples that follow,
${\cal C}$ is a submanifold of ${\cal Q}$. 
Hence the restriction of $f_{\cal Q}$ to ${\cal C}$ gives rise to 
a well--defined classical function $f_{\cal C}:{\cal C}\to {\bf R}$ 
whose quantisation reproduces the quantum observable $F$.

\section{Examples of different classical limits}\label{exam}

In the following we give some examples of the {\it dequantisation} 
of the nonrelativistic quantum mechanics of $n$ degrees of freedom. We will 
concentrate on some specific nonlinear choices for the manifold ${\cal C}$, 
namely  complex projective spaces ${\bf CP}^n$ and complex submanifolds 
thereof. Linear classical phase spaces have been dealt with in sections 
\ref{fibover}, \ref{hori}. Coherent states on spheres have been constructed 
in ref. \cite{SPHERES}.

\subsection{The standard coherent states}\label{stan}

Points in ${\bf CP}^n$ may be specified by homogeneous coordinates 
$[w_0:\ldots:w_n]$ on ${\bf C}^{n+1}$. Alternatively, holomorphic coordinates 
on ${\bf CP}^n$ in the chart with, say, $w_0\neq 0$, are given by $z_k=w_k/w_0$, 
with $k=1,\ldots,n$. 

In order to discuss coherent states it is convenient to use homogeneous coordinates.
Then we have a K\"ahler form
\begin{equation}
\omega={\rm i}\sum_{k=0}^n{\rm d} w^k\wedge {\rm d}\bar w^k,
\label{kahfor}
\end{equation}
which we take to define a symplectic structure with invariance group $U(n+1)$. 
As we are working in homogeneous coordinates we still have to mod out by $U(1)$, 
so the true invariance group of the K\"ahler form is $G=U(n+1)/U(1)\simeq SU(n+1)$.
Let $G'\subset G$ be a maximal isotropy subgroup \cite{PERELOMOV} of the vacuum state $|0\rangle $. 
Coherent states $|\zeta\rangle $ are parametrised by points $\zeta$ in the coset space $G/G'$
\cite{PERELOMOV}. Set $n=1$ for simplicity, so ${\bf CP}^1\simeq S^2$.
Then $G'=U(1)$, and coherent states $|u\rangle$ are parametrised by points $u$ 
in the quotient space $S^2=SU(2)/U(1)$.
 
We will find it convenient to recall Berezin's quantisation 
\cite{BEREZIN} of the Riemann sphere. The Hilbert space is most easily presented 
in holomorphic coordinates $z, \bar z$, which have the advantage of being almost 
global coordinates on  $S^2$. The K\"ahler potential 
\begin{equation}
K_{S^2}(z,\bar z)={\rm log}\,(1+|z|^2)
\label{sphpot}
\end{equation}
produces an integration measure 
\begin{equation}
{\rm d}\mu(z,\bar z)={1\over 2\pi {\rm i}}{{\rm d} z\wedge {\rm d}\bar z\over (1+|z|^2)^2}.
\label{sphmea}
\end{equation}
The Hilbert space of states is the space ${\cal F}_{\hbar}(S^2)$ of holomorphic functions on
$S^2$ with finite norm, the scalar product being
\begin{equation}
\langle\psi_1|\psi_2\rangle =\Big({1\over \hbar} +1\Big)\int_{S^2}{\rm d}\mu(z,\bar
z)\,(1+|z|^2)^{-1/\hbar}\,{\overline
\psi_1(z)}\,\psi_2(z).
\label{sphscalar}
\end{equation} 
It turns out that $\hbar^{-1}$ must be an integer. For $\psi$ to have finite
norm, it must be a polynomial of degree less than $\hbar^{-1}$. In fact, setting $\hbar^{-1}=2j+2$, 
${\cal F}_{\hbar}(S^2)$ is the representation space for the spin--$j$ representation of $SU(2)$,
which is the isometry group of $S^2$. 
The semiclassical regime corresponds to $j\to\infty$. 

\subsection{Trivial fibre bundles: global coherent states}\label{triexam}

Let ${\cal Q}$ be the manifold of rays in ${\cal H}$.
We define an action of the group of unitary operators $U({\cal H})$
on ${\cal Q}$ as follows: first lift ${\cal Q}$ to ${\cal H}$, 
then apply a $U({\cal H})$ transformation. In this way 
we obtain a fibre bundle whose base is ${\cal C}={\cal Q}/U({\cal H})$.
Now any two points in ${\cal Q}$ can always be connected by means of a 
transformation in $U({\cal H})$, so this ${\cal C}$ reduces to a point. 
This is an instance of the situation mentioned in section \ref{claslim}, 
that {\it not} every bundle will give rise to a reasonable classical limit.

A sensible classical limit is the following. ${\cal H}$ being infinite--dimensional, 
we may require the action of $U({\cal H})$ to act as the identity along, say, the 
first $n+1$ complex dimensions of ${\cal H}$, while allowing it to act nontrivially on 
the rest. In this way the resulting ${\cal C}={\cal Q}/U({\cal H})$ is the complex 
$n$--dimensional projective space ${\bf CP}^n$. It is the base of a 
principal fibre bundle whose total space is ${\cal Q}$ and whose fibre is $U({\cal H})$.
This bundle is trivial by construction. Triviality may also be proved 
recalling that, when the structure group is contractible, the bundle is 
automatically trivial \cite{STEENROD}. Now $U({\cal H})$ is contractible \cite{KUIPER} 
(see also section \ref{mathtech}), so all principal $U({\cal H})$--bundles 
are trivial.

Next consider the trivial vector bundle, with fibre ${\cal H}$,
that is associated with this trivial principal bundle. Triviality 
implies that one has the globally defined diffeomorphism 
${\cal Q}\simeq {\cal C}\times {\cal H}$.
Now coherent states lie on sections of this bundle.  
Hence the triviality of this bundle ensures that these coherent states 
are globally defined on ${\cal C}$. An equivalent phrasing of this statement 
is to say that the semiclassical regime is globally defined on ${\cal C}$. 
Upon quantisation, all observers on ${\cal C}$ will agree on 
what is a {\it semiclassical} vs. what is a {\it strong quantum} effect. 
Setting $n=1$ for simplicity, if one observer on ${\cal C}$ measures 
$j<\infty$, then so will all other observers. If the measure 
is $j\to\infty$, then so will it be for all other observers, too.

Now $U({\cal H})$ is the invariance group of the K\"ahler form on ${\cal Q}$
\begin{equation}
\omega={\rm i}\sum_{k=0}^{\infty}{\rm d} w^k\wedge {\rm d}\bar w^k.
\label{kahforinf}
\end{equation}
The K\"ahler form on the resulting ${\bf CP}^n$ is given in eqn. (\ref{kahfor}), 
{\it i.e.}, it is the one obtained by quotienting (\ref{kahforinf}) with this group action.
Incidentally, the metric $g$ on ${\cal Q}$ also descends to the quotient ${\bf CP}^n$,
and we can now apply Berezin's quantisation \cite{BEREZIN}. In fact we have
picked our group action precisely so as to obtain a {\it dequantisation} 
of ${\cal Q}$ to ${\bf CP}^n$ that exactly reproduces the standard clasical limit 
$\hbar \to 0$ for ${\bf CP}^n$.  Similarly, the corresponding coherent--state quantisation
\cite{PERELOMOV} is the one summarised in section \ref{stan}. This example also 
illustrates the power of fibrating ${\cal Q}$ by means of a group action. 
Yet another choice for the group action will lead to another different {\it dequantisation}.
       
\subsection{Nontrivial fibre bundles: nonglobal coherent states }\label{quatriexam}

Let us consider the Hopf bundle
\begin{equation}
S^{2n+1}/U(1)\simeq {\bf CP}^n,
\label{hopf}
\end{equation}
where the sphere $S^{2n+1}$ is the submanifold of ${\bf C}^{n+1}$ defined by
\begin{equation}
|z_0|^2 + \ldots + |z_n|^2=1,
\label{sphere}
\end{equation}
and the $U(1)$ action is 
\begin{equation}
(z_0,\ldots, z_n)\mapsto {\rm e}^{{\rm i}\alpha}\,(z_0,\ldots, z_n).
\label{maphopf}
\end{equation}
This fibre bundle is nontrivial \cite{TOPPHYS} (it describes a magnetic monopole 
of nonzero charge \cite{QFTT}). 

Let us consider the group $U(\infty)$ 
\begin{equation}
U(\infty)=\lim_{n\to\infty} U(n),
\label{uinfty}
\end{equation}
which is not to be confused with the group $U({\cal H})$ of section 
\ref{triexam}. Elements of $U(\infty)$ are $n\times n$ unitary matrices 
$u$ in any dimension $n$. In order to let them act on ${\cal H}$, 
which is infinite dimensional, we may think of $u$ as being tensored with 
an infinite--dimensional identity matrix, $u\otimes {\bf 1}$. 
Therefore  $U(\infty)$ is a subgroup of $U({\cal H})$. As was the case with
$U({\cal H})$, any two points in ${\cal Q}$ are always connected by means 
of an $U(\infty)$--transformation.

Now we have the fundamental group (see section \ref{mathtech})
\begin{equation}
\pi_1\left(U(\infty)\right)={\bf Z},
\label{homotop}
\end{equation}
so $U(\infty)$ is {\it not} contractible to a point. ${\bf CP}^n$ is also noncontractible. 
We conclude that principal $U(\infty)$ bundles over ${\bf CP}^n$ may be nontrivial 
\cite{STEENROD}. 

We define an action of $U(\infty)$ on ${\cal Q}$ as follows: first lift ${\cal Q}$ 
to the infinite--dimensional sphere $S^{\infty}$, then embed $S^{\infty}$ into ${\cal H}$ 
using equation (\ref{sphere}) in the limit $n\to\infty$, then apply a $U(\infty)$ transformation. 
We require that this action be given by eqn. (\ref{maphopf}) on the first $n+1$ dimensions 
of ${\cal H}$, {\it i.e.}, only a $U(1)$ subgroup of $U(\infty)$ will act on them. 
Along the remaining infinite dimensions we let $U(\infty)$ act unconstrained. 
In this way we obtain a principal $U(\infty)$ fibre bundle whose base ${\cal C}$ 
is ${\bf CP}^n$ and whose total space is ${\cal Q}$. This ${\bf CP}^n$ 
inherits its symplectic structure (\ref{kahfor}) by quotienting (\ref{kahforinf}) 
with the group action, so its standard quantisation reproduces the original 
quantum theory on ${\cal Q}$, up to an important difference.
Coherent states (regarded as sections of the associated 
vector bundle whose fibre is ${\cal H}$) are no longer globally defined on ${\bf CP}^n$ 
because this bundle is nontrivial by construction, and therefore it admits no global section. 

The physical implications of the local character of these coherent states
are easy to interpret. Again set $n=1$ for simplicity. In the case of the trivial 
bundle of section \ref{triexam}, the cross section of coherent states above {\it any}
observer on the base ${\bf CP}^1$ was globally defined. Hence the semiclassical 
regime was universally defined for all observers on ${\bf CP}^1$. On the 
contrary, the nontriviality of the bundle considered here 
implies that the semiclassical regime is defined only locally, and it 
{\it cannot} be extended globally over ${\bf CP}^1$. What to one observer  
appears to be a semiclassical effect  need {\it not} appear so to a different observer. 

For illustrative purposes we have explicitly constructed one particular nontrivial bundle. 
It should not be difficult to construct other nontrivial bundles such that, 
{\it e.g.}, one observer actually perceives as strong quantum ($j<\infty$) 
the same effect that another observer calls semiclassical ($j\to\infty$). 

\subsection{Submanifolds of complex projective space}\label{compsub}

Any smooth, complex algebraic
manifold $M$ given by a system of polynomial equations in ${\bf CP}^n$ has a 
natural symplectic structure \cite{ARNOLD}. Let $\iota :M\rightarrow {\bf 
CP}^n$ be an embedding of the complex manifold $M$ into complex projective 
space. Then the symplectic form $\omega$ on ${\bf CP}^n$ can be pulled 
back to a symplectic form $\iota ^*\omega$ on $M$. The fibre bundles of sections 
\ref{triexam}, \ref{quatriexam}, when pulled back to $M$,
naturally suggest new instances of classical limits. The submanifold $M$
must satisfy the integrality conditions \cite{WOODHOUSE}.

\section{Physical discussion}\label{discu}

The deep link existing between classical and quantum mechanics has been known 
for long. Perhaps its simplest manifestation is that of coherent states.
More recent is the notion that not all quantum theories 
arise from quantising a classical system. Furthermore, a given quantum model may 
possess more than just one classical limit. These facts find strong evidence 
in string duality and M--theory.

The geometric formulation of standard quantum mechanics  presented in ref. \cite{ASHTEKAR}
naturally suggests a procedure by which the passage to a classical limit may be 
performed in more than one different way. We believe this may provide a clue 
towards solving some of the conceptual problems mentioned in section \ref{intro}.

We would like to point out that we do not propose a new approach to 
quantum mechanics, nor do we cast a doubt on its conceptual framework. 
On the contrary, we stand by its standard textbook interpretation as presented, 
{\it e.g.}, in refs. \cite{GALINDO, DR}. 
Using the geometric formulation of standard quantum mechanics given in ref. \cite{ASHTEKAR},
we have simply observed that what is usually called {\it the} classical limit 
in fact corresponds to a very specific choice of a fibre bundle whose total 
space is the quantum phase space. This, in turn, univocally determines the 
classical phase space to be the expected one.  Historically the opinion has 
prevailed that {\it the classical limit is always uniquely and globally defined}. 
However, as hinted at in ref. \cite{VAFA}, we believe this latter statement must be revised 
in the light of recent developments. In fact, nowhere in the axiomatics of standard quantum 
mechanics is such a statement to be found; it probably has its origins in the chronological 
order of developments in theoretical physics. Removing the statement that 
{\it the classical limit is always uniquely and globally defined} alters neither the foundations 
nor the standard interpretation of quantum mechanics. 

The standard definition of classical limit is $\hbar\to 0$.
However, the notion of duality suggests enlarging this definition 
in order to cover other cases that, on first sight, do not fall into that category.
One possible generalisation of such a definition is the one we have 
considered here, namely, acting on the quantum phase space ${\cal Q}$ by 
means of a group $G$, so as to obtain a new phase space ${\cal Q}/G={\cal C}$. 
Calling the latter {\it classical} is justified if ${\cal C}$ is a 
symplectic manifold whose quantisation gives back the original 
quantum theory on ${\cal Q}$. If that is the case, then ${\cal C}$ truly 
is a classical limit, even if we did not arrive at ${\cal C}$ by letting 
$\hbar \to 0$.

We have emphasised the key role played by the symplectic structure in switching 
back and forth between ${\cal Q}$ and ${\cal C}$. On the contrary, the role played 
by the Riemannian metric $g_{\cal C}$ has been reduced to 
that of providing quantum numbers once a certain classical limit has 
been fixed. It is precisely through lifting the metric dependence that we
have succeeded in obtaining different classical limits for a given quantum 
theory. In this sense, as suggested in ref. \cite{NOS}, implementing 
duality transformations in quantum mechanics is very reminiscent of 
topological field theory. 

Lifting the metric dependence in favour of diffeomorphism invariance,
as in topological theories, is also important for the following reasons.
We have made no reference to coupling constants or potentials, with the understanding 
that the Hamilton--Jacobi method has already placed us, by means of suitable coordinate 
transformations, in a coordinate system where all interactions vanish. At least under 
the standard notions \cite{GALINDO} of {\it classical} vs. {\it quantum}, 
this is certainly always possible at the classical level \cite{DR}. At the quantum level, 
the approach of ref. \cite{MATONE}, which contains standard quantum mechanics 
as a limiting case, rests precisely on the possibility of transforming between 
any two states by means of diffeomorphisms. Diffeomorphism invariance is  
a very powerful tool. It can be used \cite{MATONE} in the passage from {\it classical} 
to {\it quantum}. It can also be applied in the passage 
from {\it quantum} to {\it quantum}, as in ref. \cite{MOURAO}, where Hamiltonian quantum 
theories are constructed from functional integrals in the Osterwalder--Schrader framework
\cite{OS, GJ}. The viewpoint advocated here is that it can also 
be successfully applied in the passage from {\it quantum} to {\it classical}.
 
Then the only truly {\it quantum} ingredient we have at hand is $\hbar$. In fact
one can think of quantisation, especially of deformation quantisation 
\cite{STERNHEIMER, ZACHOS}, as 
performing an infinite expansion in powers of $\hbar$ around a classical theory. 
This full infinite expansion gives the full quantum theory. {\it Dequantisation} may then be 
interpreted as the truncation of this infinite expansion to a given finite order.
As we have argued, if the quantum fibre bundle ${\cal Q}\rightarrow {\cal C}$ is nontrivial, 
this expansion in powers of $\hbar$ is local instead of global, so the 
notion of {\it classical} vs. {\it quantum} may not be globally defined for 
all observers.

\section{Mathematical discussion}\label{mathtech}

To conclude we would like to comment on some interesting mathematical 
points of our construction.

The following theorem holds \cite{STEENROD}: a sufficient condition for a fibre bundle 
to be trivial is that either the stucture group or the base manifold be contractible 
to a point. Hence the classical limit may be nonglobal only if both the structure group 
and the base manifold are noncontractible. Concerning the uniqueness of the classical limit, 
one can in principle fibrate ${\cal Q}$ in many different ways, according to the symmetries 
of the problem. 

We need to act with infinite--dimensional groups $G$ on ${\cal Q}$ in order to 
obtain a finite--dimensional quotient ${\cal Q}/G$ as a classical phase space.
When working with infinite--dimensional groups, the issue of contractibility 
deserves some care. Indeed one may topologise the group $U({\cal H})$ with 
different, nonequivalent topologies, so the contractibility of $U({\cal H})$ 
may depend on what topology one chooses for $U({\cal H})$. Two popular 
choices are the norm topology and the strong operator topology \cite{DIXMIER}. 
It turns out that both of them render $U({\cal H})$ contractible 
\cite{KUIPER, DIXMIER}. 

Concerning $U(\infty)$, the best way to topologise it is the following.
Enlarge  an $n\times n$ unitary matrix to an $(n+1)\times (n+1)$ unitary 
matrix by adding one row and one column. The group $U(\infty)$ is defined 
by performing this enlargement infinitely many times. Now the direct limit 
topology \cite{BOURBAKI} renders every matrix inclusion $U(n)\subset U(\infty)$ 
continuous, and it is the {\it maximal} topology that enjoys this property. 
Moreover, this topology also respects the fundamental group $\pi_1\left(U(n)\right)={\bf Z}$ 
in the passage $n\to\infty$, as stated in eqn. (\ref{homotop}).

One could wonder, why not use a $U(1)$ subgroup of $U({\cal H})$ instead 
of $U(\infty)$, in order to construct a Hopf bundle in section \ref{quatriexam}? 
In fact one could do so, but at the cost of rendering the whole 
infinite--dimensional bundle over ${\cal C}$ trivial; only the finite--dimensional 
subbundle corresponding to the Hopf bundle would remain nontrivial. There would be no 
contradiction, since the triviality of a given bundle does not prevent 
the existence of nontrivial subbundles. For example, given any vector bundle 
$E\rightarrow {\cal C}$ over a (compact and Hausdorff) base manifold ${\cal C}$, 
there always exists a complementary vector bundle $F\rightarrow {\cal C}$ 
such that $E\oplus F$ is trivial \cite{ATIYAH}. 

However, the situation just described is precisely what we want to avoid. 
We need the complete, infinite--dimensional bundle over ${\cal C}$ to be nontrivial 
in order for the classical limit not to be globally defined; a 
finite--dimensional subbundle will not suffice. In retrospective, this argument
also justifies our choice of $U(\infty)$ in section \ref{quatriexam}. The 
topologies considered above on $U({\cal H})$, while rendering every inclusion $U(n)\subset 
U({\cal H})$ continuous, are not the {\it maximal} topology enjoying that 
property. On the contrary, the direct limit topology on $U(\infty)$ is the maximal 
one with that property. This ensures that the addition of an infinite 
number of (spectator) dimensions to the $n$--dimensional Hopf bundle (\ref{hopf}) 
does {\it not} render the complete infinite--dimensional bundle trivial,
as would be the case with $U({\cal H})$.

Quantum--mechanical symmetries are usually implemented by the action of 
unitary operators on ${\cal H}$. The group $U({\cal H})$ thus arises naturally 
in this setup. However, any principal bundle with structure group 
$U({\cal H})$ is necessarily trivial. In retrospective, this explains 
why the classical limit is always considered to be globally defined. 
In order to bypass this difficulty we have considered the subgroup 
$U(\infty)\subset U({\cal H})$ and endowed it with a topology of its own 
(the direct limit topology) that is different from the induced topology 
it would inherit from $U({\cal H})$. Only so do we have a chance of 
rendering $U(\infty)$--bundles nontrivial. It is interesting to observe that
$U(\infty)$, instead of $U({\cal H})$, is the right group that contains all $U(n)$ groups, 
in a way that naturally respects their topologies. $U(n)$ groups arise 
naturally in theories with solitons and instantons.
In supersymmetric Yang--Mills theories and superstring theory, solitons and instantons 
lie at the heart of the notion of duality. This supports the notion that implementing 
duality transformations in quantum mechanics is in fact possible through mechanisms 
such as the one proposed here. It would also be very interesting to extend our mechanism 
to more general quantum--mechanical structures such as rigged Hilbert spaces \cite{GADELLA}.

{\bf Acknowledgments}

This work has been supported by a PPARC fellowship under grant no. 
PPA/G/O/2000/00469.


\begin{thebibliography}{99}

\bibitem{ASHTEKAR}
A. Ashtekar and T. Schilling, {\tt gr-qc/9706069};
A. Ashtekar, {\tt qr-qc/9901023}.

\bibitem{ARNOLD}
V. Arnold, {\it Mathematical Methods of Classical Mechanics}, Springer, Berlin (1989).

\bibitem{RATIU}
J. Marsden and T. Ratiu, {\it Introduction to Mechanics and Symmetry}, Springer, Berlin (1999).

\bibitem{SCHWARZ}     
J. Schwarz, {\it Nucl. Phys. Proc. Suppl.} {\bf 55B} (1997) 1;
{\tt hep-th/0007130}; {\tt hep-ex/0008017}.

\bibitem{VAFA}
C. Vafa, {\tt hep-th/9702201}.

\bibitem{KAKU}
M. Kaku, {\it Introduction to Superstrings and M--Theory}, Springer, 
Berlin (1999); {\it Strings, Conformal Fields and M--Theory}, Springer, 
Berlin (2000).

\bibitem{KOSTANT}
B. Kostant, {\it Lectures in Modern Analysis and Applications III}, ed. C. Taam, 
Springer, Berlin (1970);
{\it Group Representations in Mathematics and Physics}, ed.  V. Bargmann,  Springer, Berlin (1970).

\bibitem{SOURIAU}
J.-M. Souriau, {\it Group Theoretical Methods in Physics}, eds.  A. Janner, T. Janssen and M.
Boon, Springer, Berlin (1976);
{\it Structure of Dynamical Systems}, Birkh\"auser, Basel (1997).

\bibitem{SNIA}
\`Sniatycki, {\it Geometric Quantization and Quantum Mechanics}, 
Springer, Berlin (1980).

\bibitem{WOODHOUSE}
N. Woodhouse, {\it Geometric Quantization}, Oxford University Press, Oxford (1991).

\bibitem{BEREZIN} 
F. Berezin, {\it Sov. Math. Izv.} {\bf 38} (1974) 1116; {\it Sov. Math. Izv.} {\bf 39} (1975) 363;
{\it Comm. Math. Phys.} {\bf 40} (1975) 153;
{\it Comm. Math. Phys.} {\bf 63} (1978) 131.

\bibitem{SCHLICHENMAIER}
M. Schlichenmaier, {\tt q-alg/9601016}; {\tt q-alg/9611022};
{\tt math.QA/9902066}; {\tt math.QA/9910137}.

\bibitem{RT}
N. Reshetikhin and L. Takhtajan, {\tt math.QA/9907171}.

\bibitem{KONTSEVICH}
M. Kontsevich, {\tt q-alg/9709040};  {\it Lett. Math. Phys.} {\bf 48} (1999) 35.

\bibitem{STERNHEIMER}
For a review see D. Sternheimer, {\tt math.QA/9809056}.

\bibitem{ZACHOS}
C. Zachos, {\tt hep-th/0110114}.

\bibitem{CATFELDER}
A. Cattaneo and G. Felder, 
{\it Comm. Math. Phys.} {\bf 212} (2000) 591.

\bibitem{GALINDO}
A. Galindo and P. Pascual, {\it Quantum Mechanics}, vols. I, II, Springer, Berlin (1990).

\bibitem{COHST}   
J. Klauder and B.--S. Skagerstam, {\it Coherent States}, World Scientific, Singapore (1985);
R. Gilmore, {\it Ann. Phys. (N.Y.)} {\bf 74} (1972) 391.

\bibitem{PERELOMOV}
A. Perelomov, {\it Generelized Coherent States and their Applications}, Springer, Berlin (1986). 

\bibitem{WEINSTEIN}
J. Marsden and A. Weinstein, {\it Rep. Math. Phys.} {\bf 5} (1974) 121.

\bibitem{GS}
V. Guillemin and S. Sternberg, {\it Symplectic Techniques in Physics}, 
Cambridge University Press, Cambridge (1984).

\bibitem{MUELLER}
O. M\"uller, {\tt math-ph/0108016}.

\bibitem{NOS}
J.M. Isidro, {\tt quant-ph/0105012}.

\bibitem{STEENROD}
N. Steenrod, {\it The Topology of Fibre Bundles}, Princeton University 
Press, Princeton (1970).

\bibitem{WEINBERG}
S. Weinberg, {\it Gravitation and Cosmology: Principles and Applications of 
the General Theory of Relativity}, Wiley, New York (1972).

\bibitem{SPHERES}
B. Hall and J. Mitchell, {\tt quant-ph/0109086}.

\bibitem{KUIPER}
N. Kuiper, {\it Topology} {\bf 3} (1965) 19.

\bibitem{TOPPHYS}
A. Schwarz, {\it Topology for Physicists}, Springer, Berlin (1994).

\bibitem{QFTT}
A. Schwarz, {\it Quantum Field Theory and Topology}, Springer, Berlin (1993).

\bibitem{DR}
W. Dittrich and M. Reuter, {\it Classical and Quantum Dynamics}, Springer, 
Berlin (1994).

\bibitem{MATONE}  
A. Faraggi and M. Matone, {\it Int. Journ. Mod. Phys.} {\bf A15} (2000) 1869;
G. Bertoldi, A. Faraggi and M. Matone, {\it Class. Quant. Grav.} {\bf 17} (2000) 3965.

\bibitem{MOURAO}
A. Ashtekar, D. Marolf, J. Mour\~ao and T. Thiemann, {\it Class. Quant. 
Grav.} {\bf 17} (2000) 4919.

\bibitem{OS}
K. Osterwalder and R. Schrader, {\it Comm. Math. Phys.} {\bf 31} (1973) 83;
{\it Comm. Math. Phys.} {\bf 42} (1975) 281.

\bibitem{GJ}
J. Glimm and A. Jaffe, {\it Quantum Physics}, Springer, Berlin (1980).

\bibitem{DIXMIER}
J. Dixmier, {\it Les $C^{\star}$--Alg\`ebres et leurs Repr\'esentations}, 
Gauthier--Villars, Paris (1969).

\bibitem{BOURBAKI}
N. Bourbaki, {\it Topologie G\'en\'erale}, Hermann, Paris (1958).

\bibitem{ATIYAH}
M. Atiyah, {\it K--Theory}, Benjamin (1967).

\bibitem{GADELLA}
R. de la Madrid, A. Bohm and M. Gadella, {\tt quant-ph/0109154}.




\end{thebibliography}
\end{document}